\documentclass[3p,times]{elsarticle}
\usepackage{mathtools}
\usepackage{todonotes} 

\begin{document}
\begin{frontmatter}

\title{A network-based strategy of price correlations for optimal cryptocurrency portfolios}

\author[1]{Ruixue Jing}
\author[1,2]{Luis E. C. Rocha}
\address[1]{Department of Economics, Ghent University, Ghent, Belgium}
\address[2]{Department of Physics and Astronomy, Ghent University, Ghent, Belgium}

\begin{abstract}
A cryptocurrency is a digital asset maintained by a decentralised system using cryptography. Investors in this emerging digital market are exploring the profitability potential of portfolios in place of single coins. Portfolios are particularly useful given that price forecasting in such a volatile market is challenging. The crypto market is a self-organised complex system where the complex inter-dependencies between the cryptocurrencies may be exploited to understand the market dynamics and build efficient portfolios. In this letter, we use network methods to identify highly decorrelated cryptocurrencies to create diversified portfolios using the Markowitz Portfolio Theory agnostic to future market behaviour. The performance of our network-based portfolios is optimal with 46 coins and superior to benchmarks up to an investment horizon of 14 days, reaching up to $1,066\%$ average expected return within 1 day, with reasonable associated risks. We also show that popular cryptocurrencies are typically not included in the optimal portfolios.  Past price correlations reduce risk and may improve the performance of crypto portfolios in comparison to methodologies based exclusively on price auto-correlations. Short-term crypto investments may be competitive to traditional high-risk investments such as the stock market or commodity market but call for caution given the high variability of prices.
\end{abstract}


\end{frontmatter}

\section{Introduction}

Cryptocurrencies are digital currencies that use cryptography to validate financial transactions. The value of a cryptocurrency in USD or other fiat currencies is defined by market demand, technology, and investor sentiment~\cite{li2017technology}. The lack of governmental regulations and the intangible nature of cryptocurrencies lead to price volatility and thus scepticism on such currencies as financial assets~\cite{jacobs2018cryptocurrencies}. Nevertheless, complex systems such as the crypto market with hundreds of cryptocurrencies are expected to contain underlying mechanisms driving the dynamics of prices, and that can be exploited to extract helpful information to understand future market behaviour~\cite{hong2022impact}. This is the case given that cryptocurrencies’ prices and sentiment are fundamentally interconnected in the market. In this letter, we hypothesise that the interdependence between cryptocurrencies fluctuates less than individual prices and thus, past correlations can be used to estimate future returns. We test our hypothesis by developing a method to find the least correlated cryptocurrencies in a given (training) period to guarantee diversity and then build an optimal portfolio to achieve positive future returns while minimising risk.

Previous research on the stock market proposes to map the correlations between asset prices into networks. The minimum spanning tree (MST) of such networks contains the least correlated stocks in its periphery in contrast to highly correlated stocks in the centre of the tree~\cite{onnela2003asset}. The least correlated assets can be selected to build optimal portfolios using the Markowitz model or modern portfolio theory (MPT)~\cite{danko2018portfolio}. MSTs were also used to develop a new stock market filtering method to show that the whole market can be simulated by this reduced version~\cite{esfahanipour2015stock}. Attempts to apply such techniques outside the traditional capital market are limited. Brauneis et al. reported that MPT portfolios could obtain higher Sharpe ratios than a single cryptocurrency~\cite{brauneis2019cryptocurrency}. Their study focused on a typical cryptocurrency like Bitcoin and used an equally-weighted cryptocurrency asset class to Markowitz's mean-variance framework to derive the risk-adjusted out-performance~\cite{brauneis2019cryptocurrency}. Another study used portfolio optimisation and the generalised auto-regressive conditional heteroscedasticity (GARCH) model with only four major cryptocurrencies to show that the Foster-Hart risk optimisation method produces a more profitable portfolio than the traditional equally weighted portfolio method~\cite{kurosaki2022cryptocurrency}. The robustness of cryptocurrency portfolios to exogenous shocks was also investigated in the context of the covid-19 pandemic. Xie et al. showed that when the economic benefits and costs of adding safe-haven cryptocurrencies (like Tether) to portfolios outperform naked portfolios, which only contain Bitcoin and portfolios including traditional safe-haven assets like gold~\cite{xie2021stablecoins}. Furthermore, cryptocurrency's prospect can possibly undermine gold's hedging ability~\cite{su2020status}.

In this letter, we combine network theory methods and the MPT to build optimal cryptocurrency portfolios by exploiting the short-term memory of the crypto market. We can outperform crypto benchmarks and estimate the optimal number of cryptocurrencies needed to maximise the average return and minimise the risk in portfolios agnostic to shocks. Finally, we show that popular cryptocurrencies with high market value are only sometimes included in the optimal portfolios.

\section{Materials and Methods}

\subsection{Modern Portfolio Theory}
\label{sec:MPT}

A fundamental issue in building investment portfolios is to optimally allocate funds across assets to maximise return and minimise risk~\cite{sulistiawan2017accrual}. Not only the number of assets but also the amount of each asset are difficult to estimate. The Modern Portfolio Theory (MPT)~\cite{fabozzi2008portfolio} provides a framework to estimate optimal amounts (i.e.\ weights) of each asset to construct diversified portfolios able to maximise returns for a given level of risk. This model is based on the variance and correlation of the assets in the portfolio. Though cryptocurrencies are not financial instruments, price correlations are crucial for portfolio optimisation and diversification. Following the MPT, the optimal weight for each cryptocurrency $i$ ($i = 1,2,...,n$) in a portfolio with $n$ coins is obtained by solving the optimisation problem:

\begin{equation}
\min_\mathbf{w} \frac{1}{2}\mathbf{w}\Sigma\mathbf{w}
\end{equation} 
subject to:

\begin{equation}
\mathbf{m}^{T}\mathbf{w}\ge \mu_{0} \;\;\; ; \mathbf{e}^{T}\mathbf{w}=1
\end{equation} 
where $\mathbf{w} = (w_{1},...,w_{i}...,w_{n})$ is a vector of weights for cryptocurrency $i$, and $\Sigma$ is the covariance matrix of the cryptocurrencies' returns. The set ${\mu_{i}} = E(r_{i})$, $\mathbf{m} = (\mu_{1},...,\mu_{i},...,\mu_{n})$ is a vector of expected returns for each cryptocurrency, $\mathbf{e}$ denotes the vector of ones, and $\mu_{0}$ the acceptable expected return.

The expected return of the portfolios can then be calculated by:

\begin{equation}
r_{\text{portfolio}}=\sum_{i=1}^{n}w_{i}r_{i}
\end{equation} 

The portfolio volatility is calculated using the covariance matrix and the above-described portfolio weights.

\begin{equation}
\sigma_{\text{portfolio}} = \sqrt{\mathbf{w}^{T}\cdot \Sigma \cdot \mathbf{w}}
\end{equation}

The software Python is used to estimate the optimal weights of the portfolio using the PyPortfolioOpt module~\footnote{https://pypi.org/project/pyportfolioopt/}.

\subsection{Network-based Portfolios}

We use a fixed period of $\Delta t$ days to calculate the correlation between pairs of cryptocurrencies. Let $P_{i,t}$ be the price (in USD) of cryptocurrency $i$ at time $t$ (taken daily at midnight), with $i = 1,2,....N$ and $t = 1,2,...,t_w$ in days. The log return is defined as $r_{i,t} = \ln \left(P_{i,t} / P_{i,t-1}\right)$ and used to obtain a stationary time-series. The Pearson correlation coefficient between the log returns of $i$ and $j$ over the period $\Delta t$ is:
\begin{equation} 
\label{eq:corr cal} \rho_{ij} = \frac{Cov(r_{i},r_{j})}{\sigma_{i}\sigma_{j}} 
\end{equation}
The correlation coefficient is transformed into a distance metric~\cite{mantegna1999hierarchical} using:

\begin{equation}
\label{eq:distance matrix} d_{ij} = 2\sqrt{(1-\rho_{ij})}
\end{equation}

A network $G(N,E)$ is defined as a set of $N$ nodes $i = \left\{1,2,3,...,N \right\}$ and a set of $E$ links connecting those nodes~\cite{Costa2011}. In our model, a node $i$ represents one cryptocurrency, and the distance correlation $d_{ij}$ between cryptocurrencies $i$ and $j$ corresponds to a weighted-link $(i,j)$ connecting the respective nodes. The distance for all pairs of cryptocurrencies is mapped to a fully connected weighted network. The Minimum Spanning Tree (MST) is a filtering method used to extract an optimal weighted acyclic structure with $N$ connected nodes. The sum of the weights of all links in the tree is less than or equal to the total link-weight of every other possible spanning tree of the same network~\cite{kruskal1956shortest, tarjan1982sensitivity}. The MST gives an optimal hierarchical structure maximising the correlation between cryptocurrencies (i.e. minimising the total distance correlation). Highly correlated coins appear in the central part of the tree, whereas those poorly correlated are in the periphery. We then calculate the network centrality $L_i$ of cryptocurrency $i$, that is the average distance between cryptocurrency $i$ and any other cryptocurrency $j$ in the MST:
\begin{equation}
L_i = \frac{\sum_{i\neq j,j=1}^{N}l(i,j)}{N-1}
\end{equation}
where $l(i,j)$ is the shortest path (in terms of weighted-links) between the respective nodes. The rank of a cryptocurrency based on this centrality measure indicates the level of its decorrelation with other coins represented in the network. The portfolio is then built by adding $n$ cryptocurrencies according to their decreasing order of importance to guarantee maximum diversification, and the weighting of each cryptocurrency is calculated using the MPT (Section~\ref{sec:MPT}).

We developed three benchmark models. The first one (RAND) is an entirely agnostic model where $n$ cryptocurrencies are chosen uniformly at random from a pool of $N$ cryptocurrencies (Section~\ref{data}). The second benchmark (TOP5) is formed by the top five cryptocurrencies with the largest average market capitalisation throughout the whole period of our data set, i.e. Bitcoin (BTC), Ethereum (ETH), Litecoin (LTC), Tether (USDT) and XRP. The third benchmark is made by only Bitcoin (BTC), which is the market leader cryptocurrency.

\subsection{Cryptocurrency Price Data}
\label{data}
The original data set consists of the daily market price of 5,450 cryptocurrencies collected from several online sources (See SI). We extracted the top 1,000 coins with the highest market cap on 22.02.2022, because most coins in the data set had relatively negligible market capitalisation. Afterwards, we extracted the coins active from 15.10.2017 to 15.10.2022 ($T_{\text{total}}=1,827$ days) and excluded those with missing prices for more than 10 days. The study period was chosen to include different price patterns and endogenous and exogenous shocks, such as the explosive price growth at the end of 2017, the subsequent market crash at the start of 2018, the covid-19 outbreak, and the following global economic slowdown. The final data set thus contains time series of prices in USD for $N=157$ cryptocurrencies (Table~\ref{Tab:currency}).

\section{Results}

\subsection{Distance Correlation Matrix}

\begin{figure}[htp]
    \centering
    \includegraphics[scale=0.5]{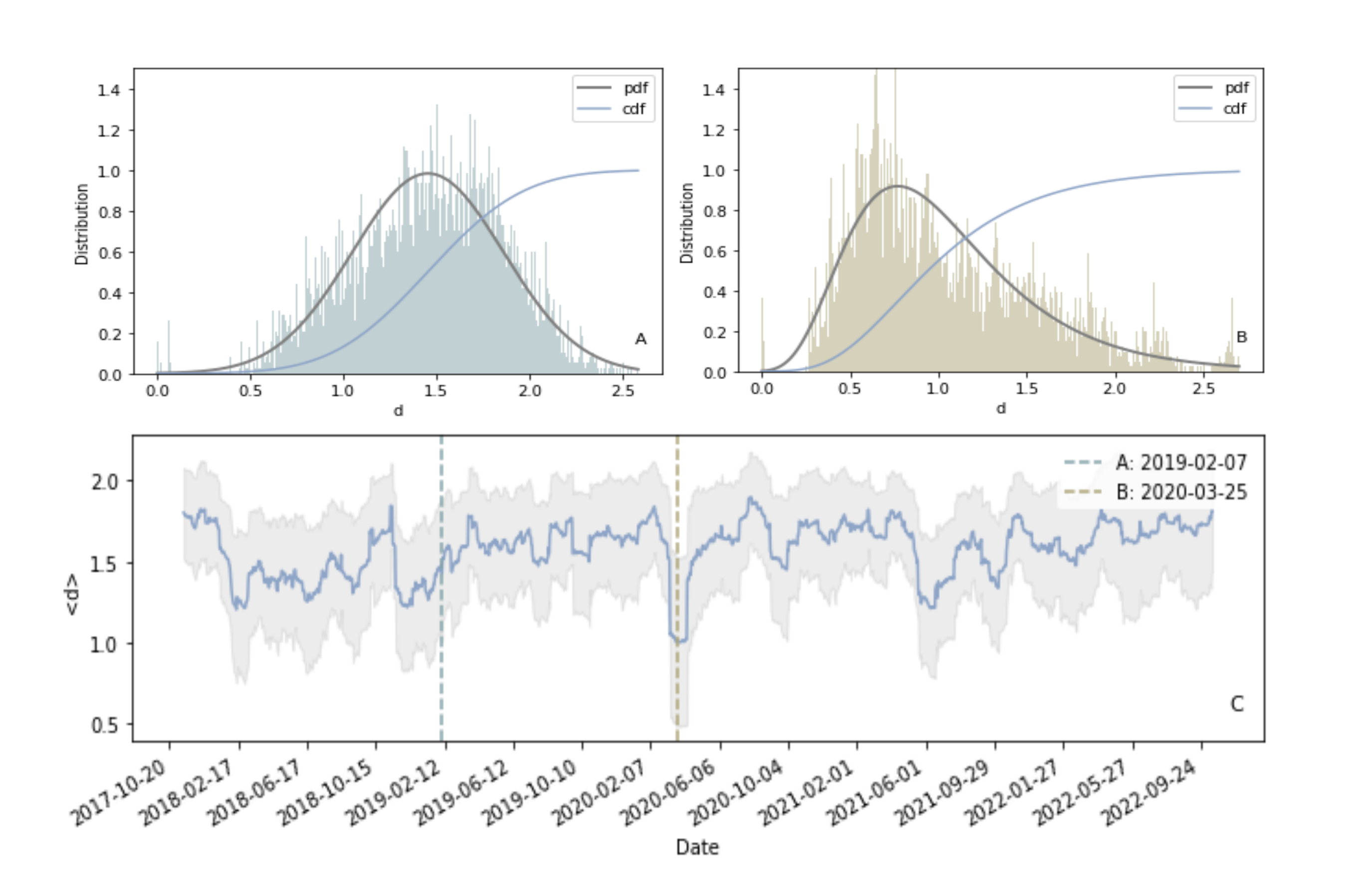}
    \caption{A) and B) show the distribution of distance correlations, respectively, during a quiet (fitted by the normal distribution, $p$-value $<.01$, $\mu = 1.45$, $\sigma = 0.41$) and a turbulent period (fitted by the log-normal distribution, $p$-value $<.01$, $\mu = -0.24$, $\sigma = 0.40$, $s = 1.19$) (See panel C) for the selected dates)). C) shows the evolution of the average distance correlation $\langle d \rangle$ during the study period and two selected dates. The shaded area shows the standard error.}
    \label{fig:dist&evolution}
\end{figure}

The interdependence between all pairs of cryptocurrencies $(i,j)$ is calculated using the distance correlation $d_{ij}(\tau)$ in intervals of $\Delta t = 30$ days shifted by $\tau=1$ day at a time (rolling window) to capture the fluctuations of the correlation networks~\cite{Rocha2017}. Figure~\ref{fig:dist&evolution}C shows that the average distance correlation $\langle d \rangle = \sum_{\tau} d_{ij}(\tau)$ strongly fluctuates between 1 and 2 indicating an overall positive correlation in the crypto market. At turbulent periods, for example, the Bitcoin crash in January 2018 and the covid-19 outbreak from February to March 2020, $\langle d \rangle$ drops significantly, indicating a strong correlation between cryptocurrencies. Figure \ref{fig:dist&evolution}B shows that pairwise correlations are highly heterogeneous (right-skewed distribution) during critical events in contrast to a normal distribution during quiet periods (Fig.\ref{fig:dist&evolution}A). While such heterogeneity may affect the quality of the portfolio, the rolling window method allows us to assess the average behaviour of the portfolio agnostic to such shocks.

\subsection{Performance Network-based Portfolio}

\begin{figure}[htp]
    \centering
    \includegraphics[scale=0.23]{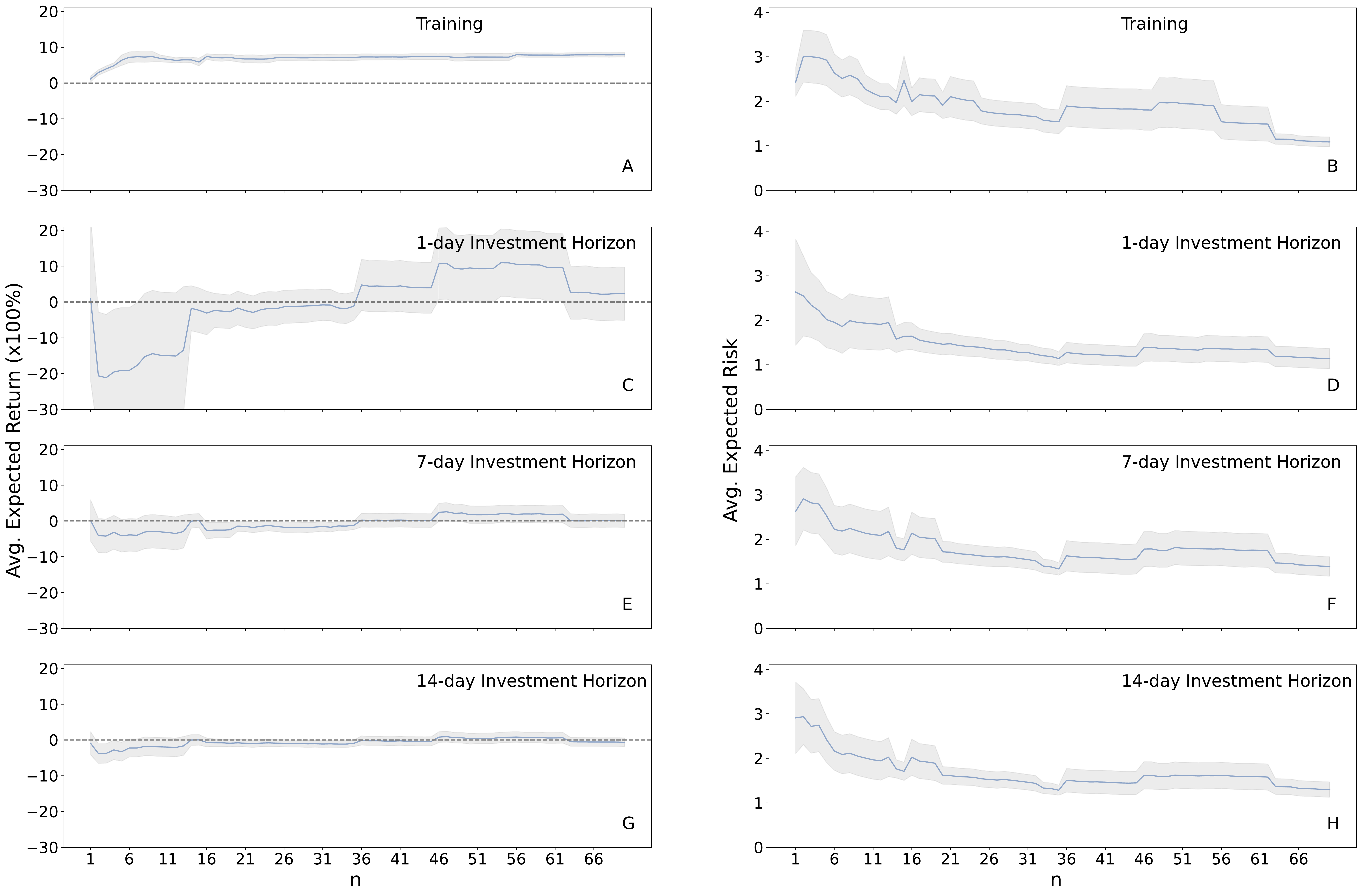}
    \caption{Average expected returns $\langle r_{\text{MST}} \rangle$ and average expected risk $\langle \sigma_{\text{MST}} \rangle$ for portfolios with $n$ cryptocurrencies during the A,B) training and C-H) future periods. The shaded area shows the standard error. The dashed vertical lines highlight the optimal number of coins for maximum return and minimum risk.}
    \label{fig:port pre 1714}
\end{figure}

The first step when designing a portfolio is to identify an optimal number of assets to achieve sufficient diversification. Too many assets (over-diversification) increases management costs, may reduce significant gains, and will not eliminate unsystematic risk~\cite{merker2019over}. To estimate the optimal number of cryptocurrencies in a purely crypto-based portfolio, we quantify the expected return and risk that our network-based portfolio model achieves for different numbers of cryptocurrencies $n$. We start with a single MST for a given time $\tau$ and select $n$ coins to the portfolio, following our proposed ranking method. Once $n$ is defined, we apply the MPT to estimate the optimal weights of each coin in the portfolio for that time $\tau$. The time interval $[\tau-\Delta t,\tau]$ is called training period. Figures~\ref{fig:port pre 1714}A,B show respectively the average expected return $\langle r_{\text{MST}} \rangle$ and average expected risk $\langle \sigma_{\text{MST}} \rangle$, where the average is taken over all times $\tau$ to include quiet and turbulent periods in the performance assessment. This first model is only based on the training period as is used as a reference portfolio. The portfolio achieves high levels of return with associated low risks; the return reaches a plateau, whereas the risk decreases for an increasing number of cryptocurrencies in the portfolio.

We now assess the performance of the same portfolio as a predictive model for future investments. We build a portfolio using data from the period $[\tau-\Delta t, \tau]$ and analyse the return and risk of that same portfolio in the period $[\tau,\tau + t_{horizon}]$. In other words, we implicitly assume that prices and correlations between cryptocurrencies remain constant. Our optimal portfolio is agnostic to future events because we only use prices during the training period. Figure~\ref{fig:port pre 1714}C shows that the portfolio performs poorly in the future (1 day) for less than $n=35$ coins but achieve high performance (highest average return) with $n=46$ coins. Figure~\ref{fig:port pre 1714}D shows that the average risk decreases by adding more coins and although it is not minimum for $n=46$, it is generally lower than for other number of coins. The portfolio performance decreases for longer time horizons, but the peaks of average return were consistently at $n=46$ (Figs.~\ref{fig:port pre 1714}E,G). The lowest average risk was achieved with $n=35$ coins (however, with slightly negative average returns). Still, they were not significantly lower than the risks for $n=46$ coins, suggesting a good compromise between return and risk.

\begin{figure}[!htp]
    \centering
    \includegraphics[scale=0.55]{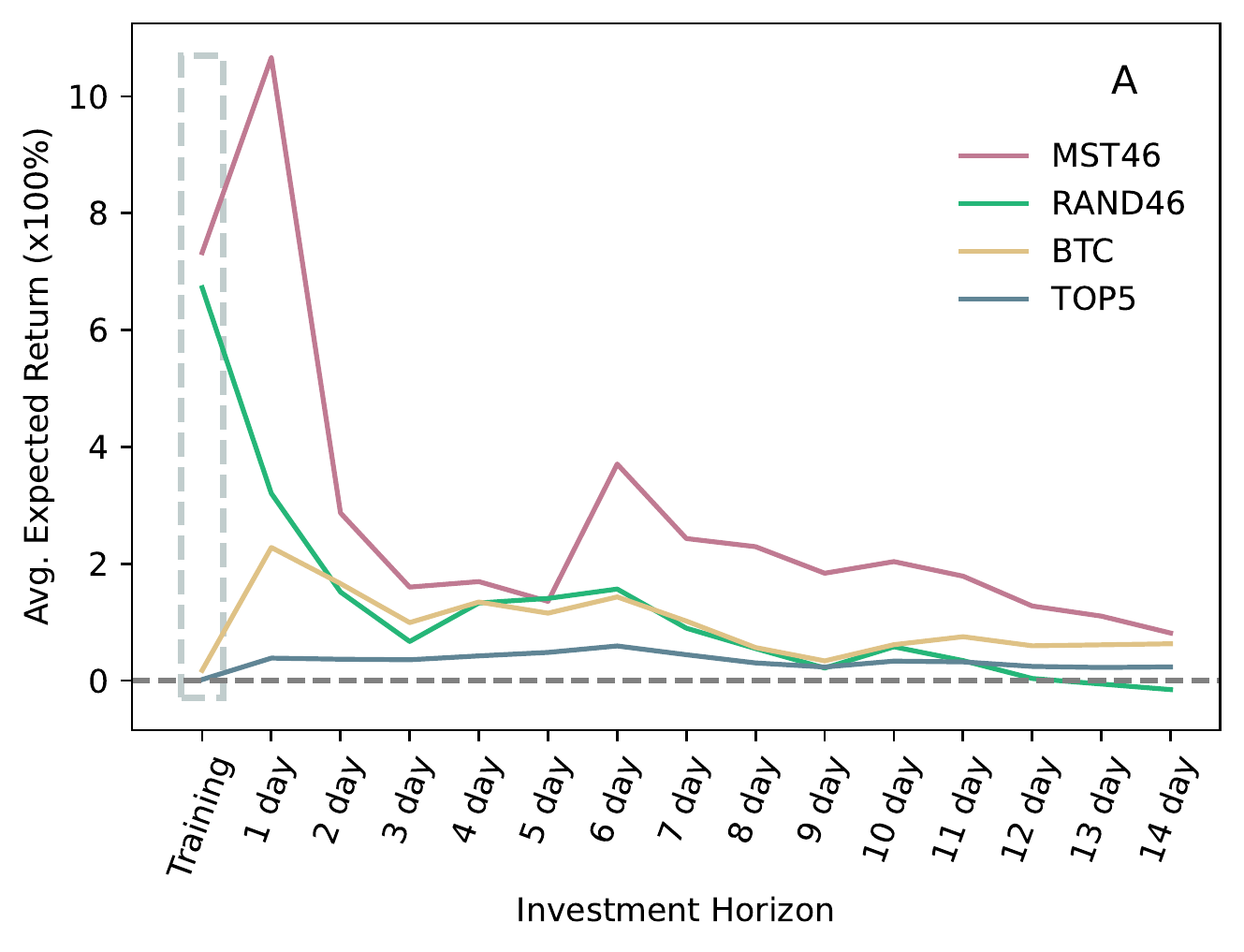}
    \includegraphics[scale=0.55]{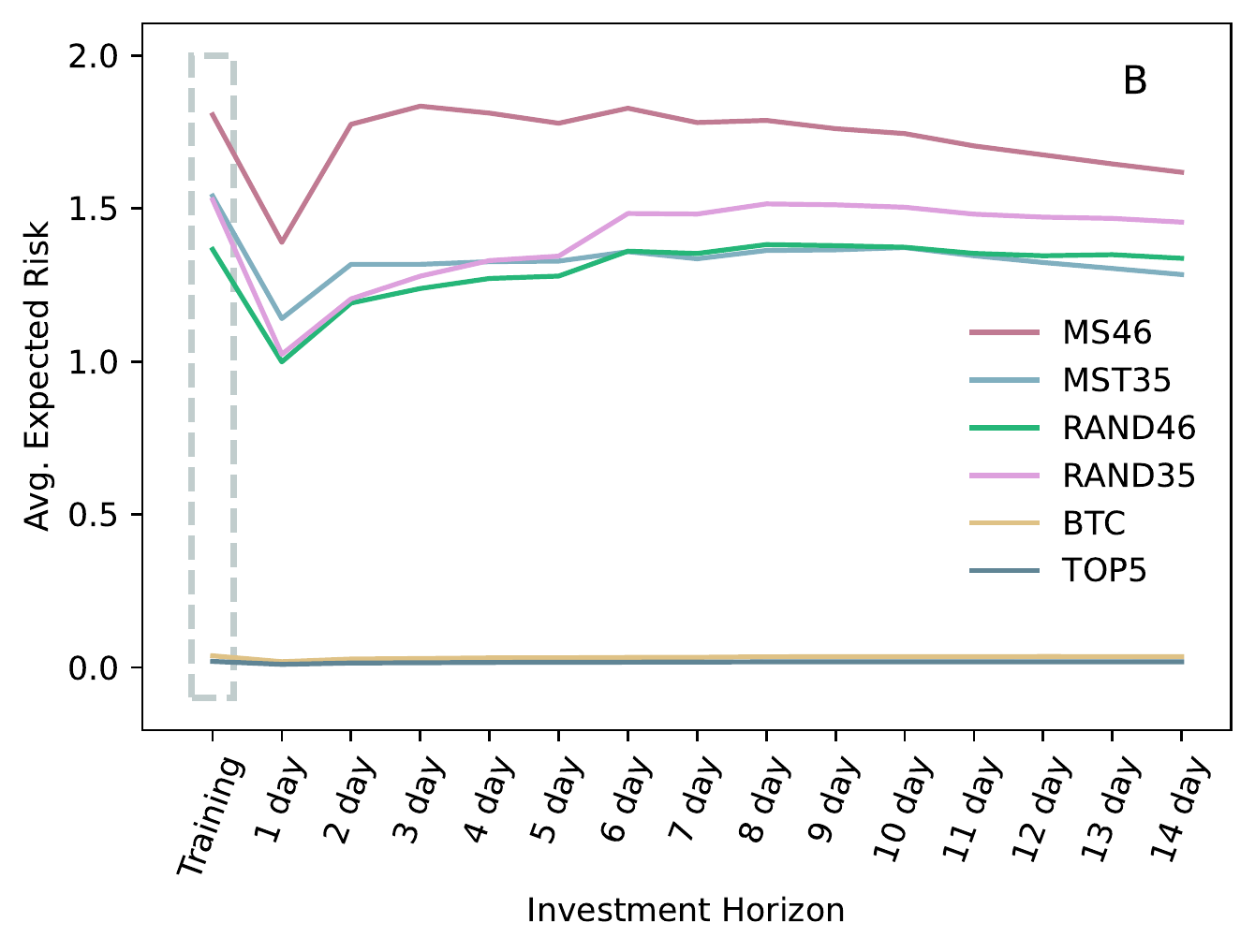}
    \caption{A) shows the average expected return and B) the average expected risk for our network-based (MST46) portfolio with $n=46$ coins and benchmark portfolios for various investment time horizons.}
    \label{fig:all cases predict}
\end{figure}

Figures~\ref{fig:all cases predict}A,B show the performance of our network-based portfolio compared to the three benchmarks in terms of average return and risk, respectively. The performance of our portfolio is always superior, achieving more than $10,000\%$ average expected return for a 1-day horizon. The peak of $\langle r_{\text{MST}} \rangle$ at 6 days suggests weekly cycles in the crypto market that require further research. The performance of a benchmark portfolio with the same number of coins chosen entirely at random (RAND46) is lower and generally equivalent to a sole bitcoin (BTC) investment. A portfolio with the top 5 coins (TOP5) gives the lowest average expected return in the two weeks. Our portfolio with $n=46$ coins consistently shows the highest average risk but not substantially higher than some other benchmarks for the 2-week investiment horizons.

\begin{figure}[!htp]
    \centering
    \includegraphics[scale=0.45]{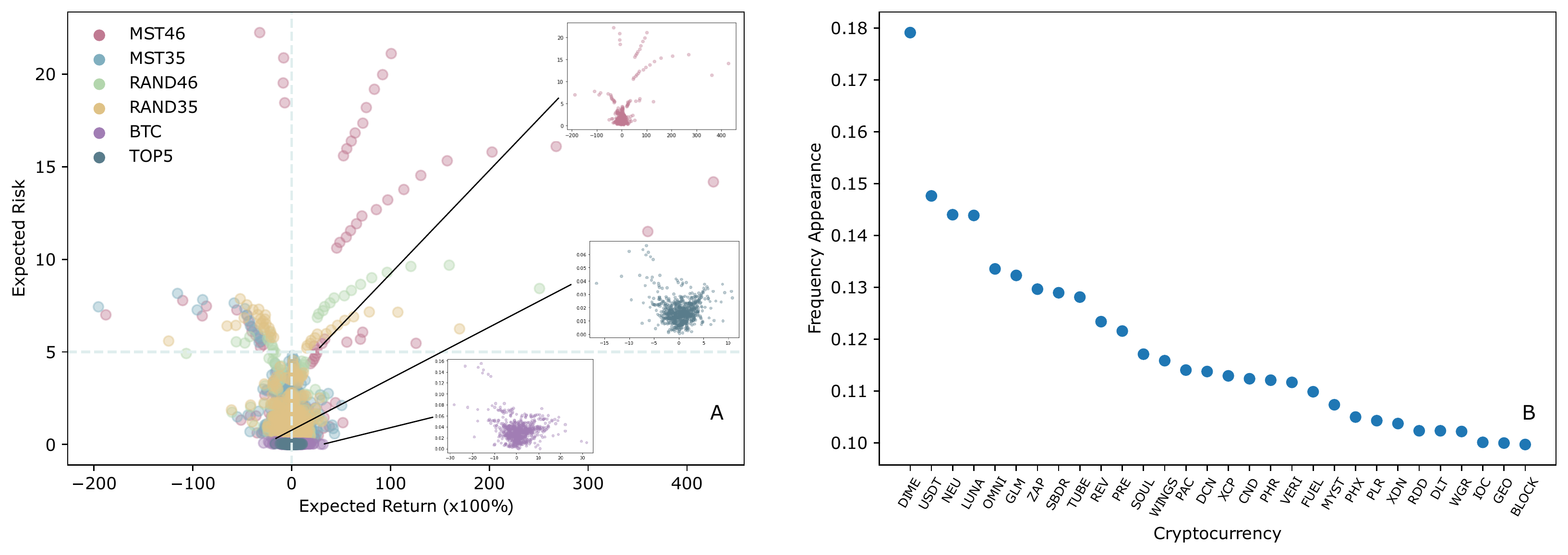}
    \caption{A) shows the expected return and risk for various portfolios during different training periods for each investment horizon (1 to 14 days, 840 points per portfolio model). Our network-based optimal portfolio (MST46) reaches the highest and second highest expected returns of $42,674\%$ and $36,036\%$ with the investment horizon of 1 day for two different training periods. The insets zoom in the performance of the portfolios for low expected return and risk. B) shows the ranking of the top 30 cryptocurrencies based on the frequency of appearance in the network-based optimal portfolios with $n=46$ coins (MST46).}
    \label{fig:Return&Risk&crypto ranking T30}
\end{figure}

Figure~\ref{fig:Return&Risk&crypto ranking T30}A shows a scatter plot of the return and risk for different portfolios (including the benchmarks) for various investment horizons. Four regions are defined by positive vs negative returns and low vs high risk. Conservative portfolios give positive returns with low risk (bottom left), but aggressive investors may consider high returns associated with increased risks (top right of graph). The benchmarks TOP5 and BTC provide the lowest risks, generating positive returns, respectively, in $59.3\%$ and $57.8\%$ of the agnostic periods (for all investment horizons); that compares to $53.2\%$ in our network-based portfolio (MST46). The relative decrease in risk is generally disproportional to the decrease in the returns. An increase in the investment horizon from 1 day to 2 days, during the training period with highest return, decreases the return of our portfolio by $37.28\%$ and the risk by $13.51\%$. The best (1-day) and the worst (14 days) investment horizons using our network-based portfolio show a relative decrease of $92.40\%$ in returns, whereas the risk decrease by only $15.83\%$. In contrast, for the benchmark TOP5, the worst investment horizon decreases the return by $62.71\%$ whereas the risk increases by $5.88\%$.

Figure~\ref{fig:Return&Risk&crypto ranking T30}B shows the top 30 most selected coins in our best-performing portfolio with $n=46$ coins (MST46) considering all training periods ($\tau$). These frequent coins appear in at least $9.97\%$ of the portfolios, with Dimecoin (DIME) included in $17.91\%$ of them. Although DIME is not among the top market-value cryptocurrencies, it has a relatively fast block time of 30 seconds (i.e. transactions can be confirmed quickly), making it useful for everyday micro-transactions. Except for Tether (USDT) ($14.8\%$), the cryptocurrencies included in our benchmarks are generally outside the top 30. Bitcoin (BTC) is used $3.3\%$ of the times, with Ethereum (ETH), Litecoin (LTC) and XRP used, respectively, $7.1\%$, $1.1\%$, and $1.8\%$. USDT is a cryptocurrency designed to maintain a stable (1:1) value relative to USD, and thus not recommended for speculative investment.

\subsection{Financial Analysis}

Although the crypto market is relatively young compared to the stock and exchange markets, it has played an increasingly important role in the financial system, calling the attention of central banks and often reflecting the bilateral relations between countries~\cite{qin2022inevitable}. Investors with different profiles have joined the crypto market, but speculative investors dominate it leading to a highly fluctuating behaviour. The MPT assumes that markets are efficient~\cite{mangram2013simplified}. The capital market efficiency hypothesis indicates that when new information comes into the market, it is immediately reflected in asset prices. Neither technical nor fundamental analysis can generate excess returns. All stocks trade at their fair market value, which means only inside information can result in outsized risk-adjusted returns~\cite{fama1970efficient}. As more cryptocurrencies are involved, and more information is revealed, the cryptocurrency market became more efficient in 2017–2019~\cite{le2020efficiency}. As in the traditional capital market, investors can only diversify the unsystematic risk associated with the corresponding cryptocurrency’s risk (e.g., technology and credibility). The inherent market (systematic) risk, which cannot be eliminated (associated for example with regulatory and macroeconomics forces), will anyways affect all coins~\cite{jalan2023systemic}, thus calling for portfolio diversification as a risk-reducing mechanism.

There is room for profit when trading in this highly volatile crypto market~\cite{balcilar2017can}, but the associated risk is non-negligible. Most investors in the capital market are conservative and rational, and they want their allocations to be profitable. Due to uncertain investment information, regular investors follow herd behaviour~\cite{lin2013investors} and mimic peer investors. Some investors also follow rational herding in which market participants react to information about the behaviour of other market agents or participants rather than the behaviour of the market per se and of fundamental transactions~\cite{devenow1996rational}. Consequently, most uncertain crypto market investors tend to buy ``popular'' coins like Bitcoin and Ethereum, leading to higher investment barriers (fear, inequitable access, and insufficient funds~\cite{lieberman1987excess}) than observed in the traditional capital market. More funds are generally less relevant in the crypto market since entry investments are low. However, our results on the frequency of specific cryptocurrencies on the optimal network-based portfolios show that the most ``popular'' coins do not lead to the most profitable portfolios. That reflects the dynamics of the crypto market where non-mainstream cryptocurrencies have technological advantages such as faster transaction time or may be designed for specific industries or niches, providing more attractiveness and thus better diversification than mainstream cryptocurrencies. On the other hand, they have lower liquidity and trading volumes, potentially increasing their risk.

The distance correlation network is a useful representation of the crypto market because it captures the price inter-dependencies between cryptocurrencies~\cite{allen2009networks}. Like the traditional capital market, the crypto market is not immune to contagion risk~\cite{da2019herding}. In our analysis, this is seen by the temporal evolution of the correlations, where correlations increase with exogenous shocks, a similar phenomenon observed in traditional markets~\cite{broadstock2012oil}. Furthermore, portfolios in traditional markets increasingly benefit (i.e. higher returns and lower risks) with diversity, and observe higher risk for short-term investment~\cite{stewart_2019}. Such portfolios may be formed by hundreds or thousands of stocks (e.g., combined holdings of the mutual fund or exchange-traded funds). On the other hand, the network-based crypto portfolios observe the best performance in the short-term with best returns (and minimum risks) with around $25\%$ of the total number of available cryptocurrencies ($N=157$).

The diversification theory states that an efficient investing technique requires allocating mixed products to offset the loss of one asset class~\cite{chatterjee1991link}. Selecting cryptocurrencies using MST in combination with the MPT to optimise the portfolio weights satisfies the diversification principle. A network-based portfolio with $n=46$ cryptocurrencies can achieve significantly high returns, whereas less diverse portfolios bring lower returns and higher risks. Investors must balance their asset portfolios based on investment goals and risk tolerance. While there is no one-size-fits-all approach to portfolio allocation, diversification in traditional capital markets usually include more than ten stocks to benefit~\cite{evans1968diversification}. This result may indicate a more general relationship where potentially an optimal number of cryptocurrencies exists. Increasing the number of assets beyond that point accumulates systematic risks to the detriment of a potential return increase. In contrast to decreasing performance in the long term, the high short-term return indicates that the market is not entirely uncorrelated. Such correlations could be exploited to improve estimators using multivariate regression~\cite{chamberlain1982multivariate} instead of univariate temporal models~\cite{murray2023forecasting}. The MST is one available technique to maximise the hierarchy of correlations between cryptocurrencies~\cite{coelho2007evolution} but other network clustering methods could be adopted~\cite{Costa2011}.

\section{Conclusion}

The crypto market is volatile and highly speculative, restricting thus opportunities for long-term investment. Our study found that although the crypto market shows highly fluctuating prices daily compared to the traditional financial market, patterns of price inter-dependencies can be exploited to create efficient agnostic portfolios. Using a network-based methodology, we designed portfolios that achieved high returns, with acceptable risks, within a short-term investment horizon but observing degrading performance in the long term. The results suggest that cross-correlations might be higher than temporal auto-correlations (i.e. absence of memory), thus opening up a venue for further research on efficient portfolio modelling. We also found that our best-performing portfolios contained 46 coins (from a pool of 157 cryptocurrencies) and barely included the most popular coins regarding market capitalisation. Some level of diversification has already been facilitated for regular investors via major crypto exchange platforms. Future research should however focus on reducing the risk associated with crypto portfolios, possibly increasing the investment horizon. Using trade indicators combined with cross-correlations might help achieve those goals. Given the non-negligible risks associated with our portfolios, cryptocurrency investment remains recommended to aggressive investors who can pool sufficient diversification in their crypto portfolios.

\section*{Acknowledgements}
The authors declare that they have no conflict of interest. R.J. is funded by the China Scholarship Council (CSC) from the Ministry of Education of P.R. China.

\bibliography{mybibfile}

\setcounter{figure}{0}
\setcounter{table}{0}
\renewcommand{\thefigure}{S\arabic{figure}}
\renewcommand{\thetable}{S\arabic{table}}

 \section*{Supplementary Information}

 \subsection{Data source}

The daily price data of the cryptocurrencies used in our study was obtained from the following websites:
www.investing.com, coinmarketcap.com, www.coindesk.com, and www.marketwatch.com. Table~\ref{Tab:currency} shows the code of all the cryptocurrencies used in our study.

\begin{table}[htp]
    \centering
    \begin{tabular}{|l|l|l|l|l|l|l|l|l|l|}
    \hline
        ADA & BLOCK & OCEANp & PPC & ETH & QTUM & STORJ & LBC & MTL & XCP \\ \hline
        ADX & BNB & OCN & CND & ETP & RCN & STRAX & LINK & MYST & XDN \\ \hline
        AE & BNT & OK & CVC & EVX & RDD & SWFTC & LRC & NAS & XEM \\ \hline
        AION & BTC & OMG & DASH & FAIR & RDN & SYNX & LSK & TOA & XLM \\ \hline
        AMB & BTM & OMNI & DCN & FCT & REV & SYS & LTC & TRX & XRP \\ \hline
        ANT & BTS & ONION & DCR & FLASH & RLC & GBYTE & LUNA & TUBE & XST \\ \hline
        AOAR & BTU & PAC & DENT & FLO & SALT & GEO & MAID & UBQ & XTZ \\ \hline
        ARDR & NEBL & PART & DGB & FTC & SBDR & GLM & MANA & USDT & XVG \\ \hline
        ARK & NEO & PASC & DIME & FUEL & SC & GRC & MDA & VAL & ZAP \\ \hline
        AST & NEU & PAY & DLT & FUN & SCRT & ICX & MDT & VERI & ZCL \\ \hline
        BAT & NLG & PHR & DNT & GAME & SMART & IGNIS & MHC & VET & ZEC \\ \hline
        BCD & NMC & PHX & DOGE & GAS & SNC & IOC & IOTA & VIA & ZEN \\ \hline
        BCH & NMR & PIVX & EDG & PPT & SNM & JNT & MKR & VIB & ZRX  \\ \hline
        BCN & NXS & PLR & EMC2 & PRE & SNT & KEY & MLN & WAVES & ~ \\ \hline
        BITCNY & NXT & PND & ENG & PRO & SOUL & KMD & MONA & WGR & ~ \\ \hline
        BLK  & OAX  & POT  & ETC  & QRL  & STEEM  & KNC  & MTH  & WINGS  & ~ \\ \hline
    \end{tabular}
    \caption{The list of the $N=157$ cryptocurrencies used in our study.}
    \label{Tab:currency}
\end{table}

\subsection{Minimum Spanning Tree}

\begin{figure}[htp]
    \centering
    \includegraphics[scale=0.3]{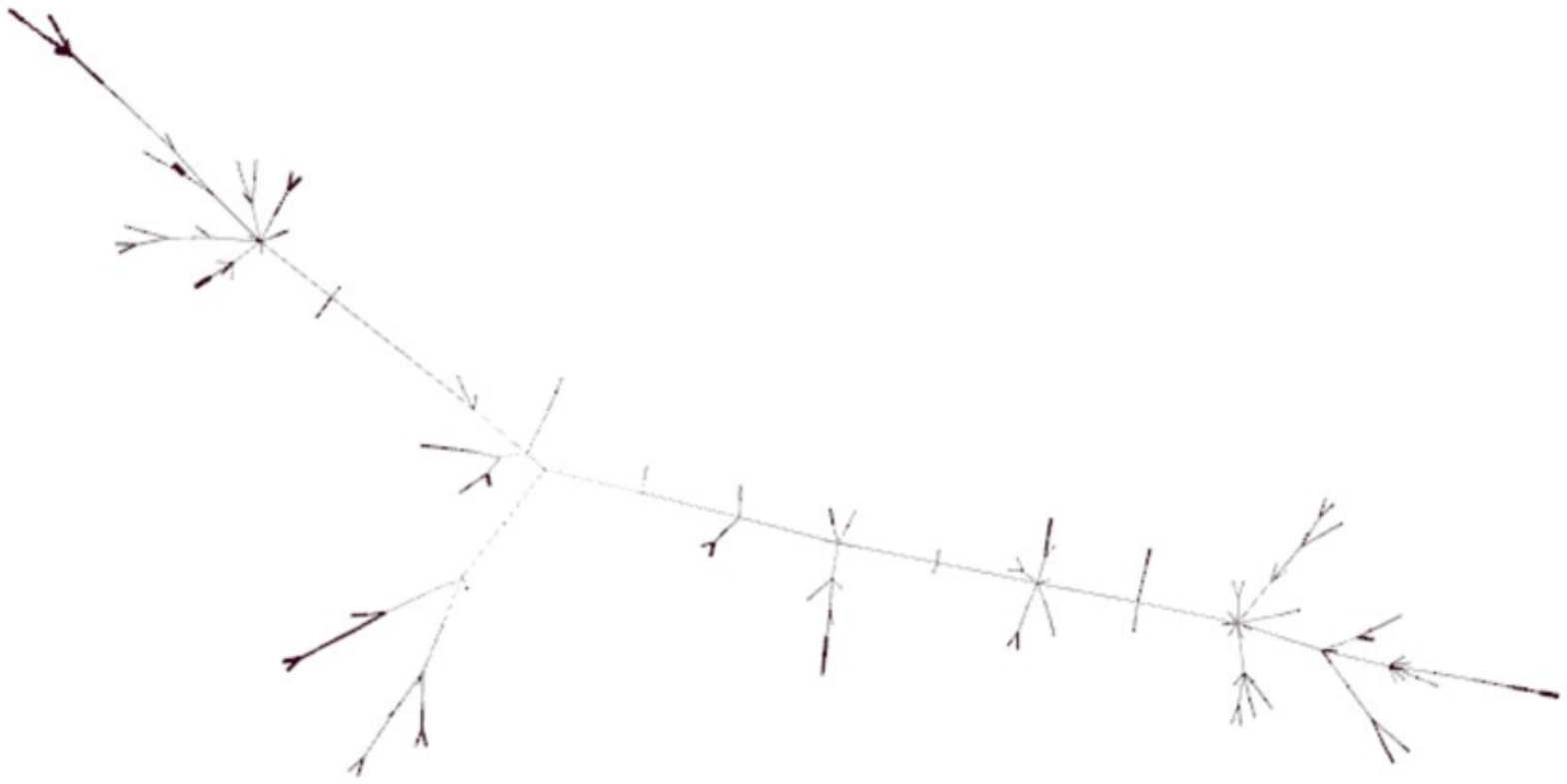}
    \caption{Minimum Spanning Tree of the crypto correlation distance network on April 15, 2018. }
    \label{fig:eg MST}
\end{figure}

Figure~\ref{fig:eg MST} shows an example of the crypto correlation distance network filtered by using the Minimum Spanning Tree method (MST) on April 15, 2018.

\subsection{Performance for various investment horizons}

\begin{figure}[htp]
    \centering
    \includegraphics[scale=0.2]{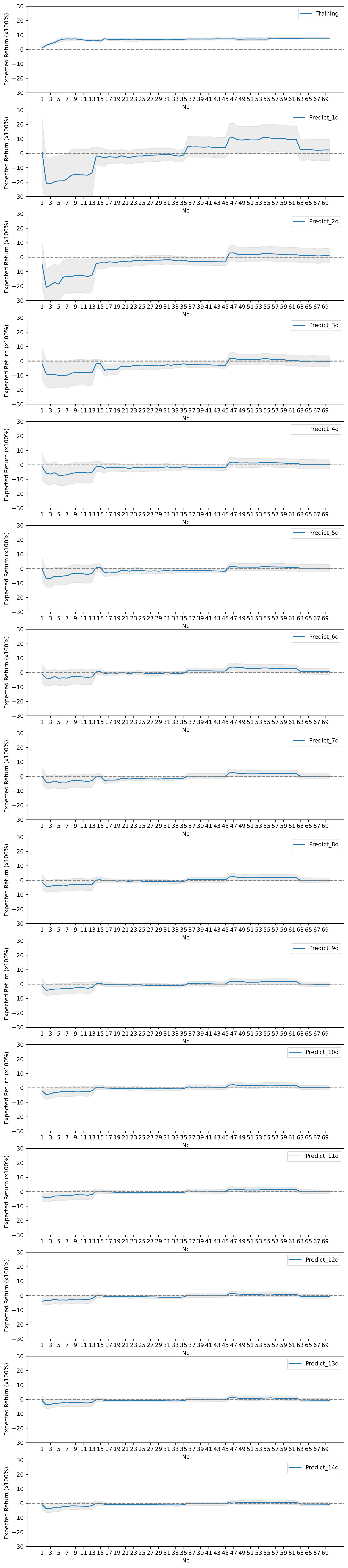}
    \includegraphics[scale=0.2]{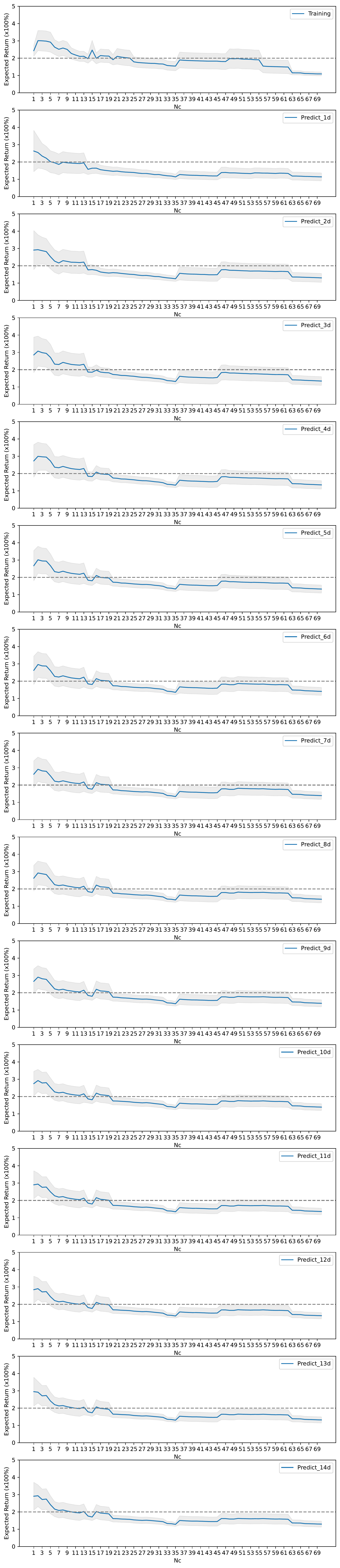}
    \caption{Predicted average expected return and average volatility for investment horizons from 1 to 14 days. Grey areas show the respective standard erros. }
    \label{fig:port pre 1-14}
\end{figure}

Figure~\ref{fig:port pre 1-14} shows the predicted average expected returns (with standard errors) and average expected volatility (risk, with standard errors) for the network-based portfolios with $n=1$ to $n=70$ cryptocurrencies.

\subsection{Frequency of occurence of cryptocurrencies in optimal portfolios}

\begin{figure}[htp]
    \centering
    \includegraphics[scale=0.23,angle=270]{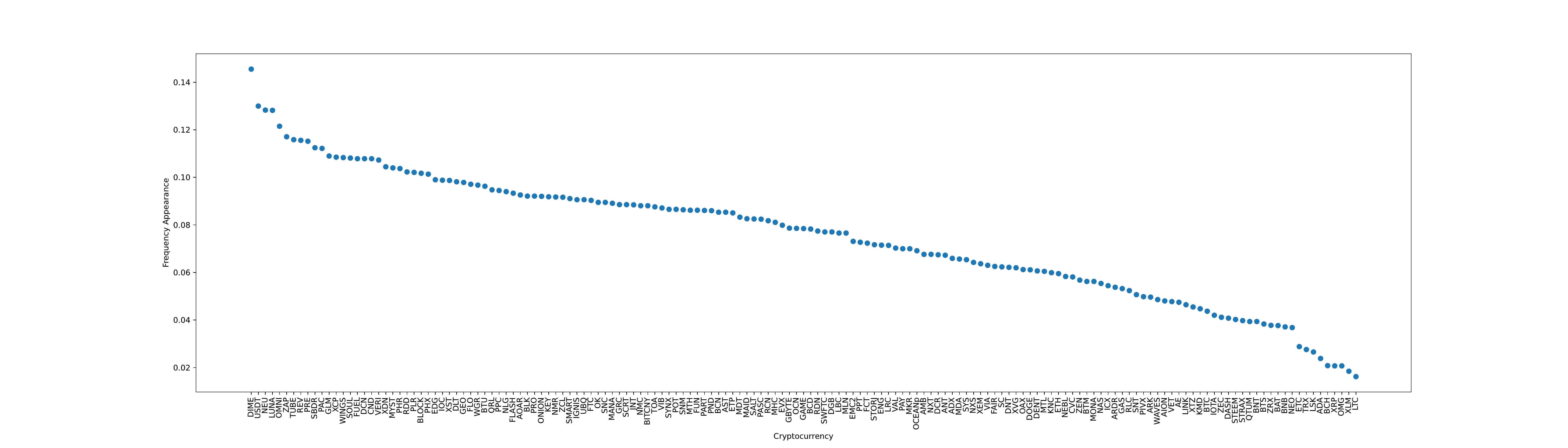}
    \caption{Ranking of the cryptocurrencies according to its frequency of use in the optimal network-based portfolios.}
    \label{fig:crypto ranking}
\end{figure}

Figure~\ref{fig:crypto ranking} shows the frequency of occurrence of cryptocurrencies used in the optimal network-based portfolios.

\end{document}